\documentclass[11pt]{article}
\usepackage[margin=1in]{geometry}
\usepackage{amssymb, amsmath, amsthm, bm, epstopdf, graphics, natbib, setspace}

\newtheorem{prop}{Proposition}[section]

\newtheorem{corollary}{Corollary}[section]
\newcommand{\argmin}{\operatornamewithlimits{argmin}}
\title{Parameter Estimation using Empirical Likelihood combined with Market Information}
\author{Steven S.G. Kou$^*$, Tony Sit$^\dagger$,and Zhiliang Ying$^\dagger$\\
\small{$^*$\textit{Department of Industrial Engineering and Operations Research, Columbia University, New York NY 10027}}\\
\small{$^\dagger$\textit{Department of Statistics, Columbia University, New York NY 10027}} \\
\footnotesize{sk75@columbia.edu \quad tony@stat.columbia.edu \quad zying@stat.columbia.edu}}
\date{}
\begin{document}
\doublespacing
\maketitle
\begin{abstract}
During the last decade L\'evy processes with jumps have received increasing popularity for modelling market behaviour for both derviative pricing and risk management purposes. \cite{Chan_etal-2009-JASA} introduced the use of empirical likelihood methods to estimate the parameters of various diffusion processes via their characteristic functions which are readily avaiable in most cases. Return series from the market are used for estimation. In addition to the return series, there are many derivatives actively traded in the market whose prices also contain information about parameters of the underlying process. This observation motivates us, in this paper, to combine the return series and the associated derivative prices observed at the market so as to provide a more refletive estimation with respect to the market movement and achieve a gain of efficiency. The usual asymptotic properties, including consistency and asymptotic normality, are established under suitable regularity conditions. Simulation and case studies are performed to demonstrate the feasibility and effectiveness of the proposed method.
\end{abstract}
\section{Introduction}
Brownian motion and normal distribution have been widely used in the Black–Scholes option-pricing framework to model the return of assets. Stylised facts, however, contradict with the model assumptions specified in the Black-Scholes framework. This motivated studies to modify the Black-Scholes model to explain the empirical phenomena. One direction of extension is to model the asset return dynamics through L\'evy processes which are able to capture jumps and the asymmetric leptokurtic features. Readers are referred to \cite{Schoutens-2003-Wiley} for a full account of L\'evy processes and their applications in finance.\\ 

Empirical likelihood, introduced by \cite{Owen-1988-Bka}, provides an alternative, nonparametric approach to inference. By placing a probability $p_j$ on the $j$th observation and computing a profile likelihood, the method can be used to construct nonparametric point estimation as well as confidence regions for the parameters of interest. \cite{Qin_Lawless-1994-AoS} first linked estimating equations with empirical likelihood. In the paper, they developed methods of combining information about parameters in over-constrained optimisation problems which are frequently discussed in econometrics or finacial literature. When the number of estimating equations is larger than the number of parameters of interest, the empirical likelihood estimation procedure will automatically combine the constaints by assigning them appropriate weights and produce an efficient estimate. \cite{Owen-2001-CRC} gives a good review on the development and applications of empirical likelihood.\\

Statistical inference based on the characteristic functions was proposed by \cite{Feuerverger_Mureika-1977-AoS}, \cite{Feuerverger_McDunnough-1981a-JRSSB} for independent observations. \cite{Chan_etal-2009-JASA} suggested using characteristic functions as constraints for the empirical likelihood estimation. Such an approach makes use of the advantage that the characteristic functions of many diffusion processes are readily available, but it does not incorporate information from the market including derivative prices, for instance, which provide informative and most-updated knowledge of the parameters of interest. The key goal of this paper is to discuss how one can make use of the market data in the empirical likelihood estimation procedure to obtain more accurate estimates.\\

Let $\{S_t\}_{t\geq0}$ be a continuous-time L\'evy process that records the evolution of a financial security over a period of time. Assuming that $S$'s are observed over a collection of discrete time points: $0, \delta, 2\delta, \ldots, n\delta$, over a time span $[0, n\delta]$, we can treat the difference of any two consecutive observations, i.e. the increments, as a set of independent observations with the same distribution since increments of a L\'evy process are independently and identically distributed. In other words
\begin{equation*}
R_j := \log S_{j\delta} - \log S_{(j-1)\delta} \sim_{iid} F_{\boldsymbol\theta}, \text{ say} 
\end{equation*}
whose characteristic function is given by
\begin{equation*}
\phi(t; \boldsymbol\theta) = E^{\mathbb{P}}[\exp\{itR_j\}; \boldsymbol\theta] = \int \exp\{itr\}F_{\boldsymbol\theta}(dr),
\end{equation*}
where $\mathbb{P}$ denotes the expectation taken under the physical measure and $\boldsymbol\theta$ denotes the parameters of interest that governs the process $\{\log S_t\}_{t \geq 0}$. Of course, using the maximum likelihood approach can produce the most efficient parameter estimates. This is, however, only possible when the density function is readily available, which is not the case for most of the L\'evy processes. In this paper, we follow \cite{Chan_etal-2009-JASA} to formulate an estimation procedure using the empirical likelihood with characteristic functions as one of the constraints. Observe that a characteristic function contains the same amount of model information as what a probability density function can carry, it is sensible to incorporate them as one of the estimating equations. Instead of having the return sequence as the only source of data, we can, in fact, incorporate information from actively-traded derivatives in order to provide a more timely estimate of the model parameters. In this paper, prices of European call options on the same underlying asset are used as moment constraints for empirical likelihood estimation procedure. Due to put-call parity between Euorpean calls and their put counterparts, it suffices to include just call prices as the put counterparts should contain the same amount of information.\\

The remainder of the paper is organized as follows: we first define the notation and describe the methodology needed in Section 2. Sections 3 provides readers with specific examples on how to apply the results in Section 2 to carry out the estimation procedure. Section 4 extends the model to multiperiod case. A simulation study and a case study are given in Section 5 and 6 respectively, followed by a discussion in Section 7. Proofs are relegated to the Appendix.
\section{Methodology}
\subsection{Single Period Model}
Throughout this section, we assume that $R_1, \ldots, R_n$ are iid random variables with distribution $F$ and the characteristic function $\phi(t; \boldsymbol{\theta})$, whose closed-form fomulation can be readily obtained. To begin, we start with the simpliest possible: In addition to the return series, we also observe risk-free rate $r$ as well as a call option with maturity $\delta$ and strike $K$. L\'evy processes have independent stationary increments and so the above set up fits L\'evy process.\\

Following \cite{Qin_Lawless-1994-AoS} and \cite{Chan_etal-2009-JASA}, we study the maximum empirical likelihood estimator (MELE) based on constraints due to both the characteristic function as well as option prices as follows. First, it is easy to see that the equation $\phi(t; \boldsymbol{\theta}) = E[e^{itR_j}]$ provides us with two constraints on $\boldsymbol{\theta}$:
\begin{eqnarray*}
\sum^n_{j=1}p_j\cos(tR_j) = Re[\phi(t; \boldsymbol{\theta})] & \text{and} &
\sum^n_{j=1}p_j\sin(tR_j) = Im[\phi(t; \boldsymbol{\theta})],
\end{eqnarray*}
where $Re(z)$ and $Im(z)$ denote respectively the real and the imaginary parts of $z$.\\

For the option constraint, denote $\widetilde{c}(S_{n\delta}, K, r, \delta, \boldsymbol{\theta})$ the call price observed at time $n\delta$, with the underlying asset price $S_{n\delta}$ and strike $K$ that matures at $(n+1)\delta$. To simplify the notation, for the rest of the paper, we suppress the subscript $\delta$ and use $S_n$ and $R_n$ to denote the underlying asset price and the associated return at time $n\delta$ respectively.\\

Observe that
\begin{eqnarray}
0 & = & E^\mathbb{Q}[e^{-r\delta}\max\{S_{n+1}-K, 0\}|S_n] - \widetilde{c}(S_n, K, r, \delta, \boldsymbol{\theta}) \nonumber\\
  & = & E^\mathbb{Q}[e^{-r\delta}\max\{S_ne^{R_{n+1}}-K, 0\}|S_n] - \widetilde{c}(S_n, K, r, \delta, \boldsymbol{\theta})\nonumber\\
  & = & E^\mathbb{P}\left[e^{-r\delta}\max\{S_ne^{R_{n+1}}-K, 0\}\frac{d\mathbb{Q}}{d\mathbb{P}}(R_{n+1};\boldsymbol{\theta}) - \widetilde{c}(S_n, K, r, \delta, \boldsymbol{\theta})\bigg|S_n\right], \label{eq:optioncon}
\end{eqnarray}
where $\frac{d\mathbb{Q}}{d\mathbb{P}}(R; \boldsymbol{\theta})$ represents the Radon-Nikodym derivative (or the density ratio) of $R$ that adjusts the difference between the probabilities defined under the physical and a risk neutral measures.\\

Option prices are usually specified through moneyness which is denoted by $m = S_n/K$. \eqref{eq:optioncon} can be rewitten as
\begin{eqnarray}
0 & = & E^\mathbb{P}\left[S_n\left(e^{-r\delta}\max\{e^{R_{n+1}}-m, 0\}\frac{d\mathbb{Q}}{d\mathbb{P}}(R_{n+1};\boldsymbol{\theta}) - \frac{\widetilde{c}(S_n, K, r, \delta, \boldsymbol{\theta})}{S_n}\right)\bigg|S_n\right]\nonumber\\
& = & E^\mathbb{P}\left[e^{-r\delta}\max\{e^R-m, 0\}\frac{d\mathbb{Q}}{d\mathbb{P}}(R;\boldsymbol{\theta}) - c(m, r, \delta, \boldsymbol\theta)\right],
\end{eqnarray}
which gives an additional constraint. Note that $c(m, r, \delta, \boldsymbol\theta) = \widetilde{c}(S_n, K, r, \delta, \boldsymbol{\theta})/S_n$ is independent of $S_n$ in most cases; see Section 3.\\

The above derivation differs from \cite{Stutzer-1996-JF}'s canonical approach in which case historical returns are used to construct $n$ possible values for the asset price one period from now, i.e. 
\begin{equation*}
S_{n+1} = S_ne^{R_j}, j = 1, \ldots, n.
\end{equation*}

That is, the previous realised returns are used to construct possible prices at $(n+1)\delta$. \cite{Walker_Haley-2010-JoFM} used a similar approach to investigate alternative tilts for non-parametric pricing. They proposed the following estimating equation: 
\begin{equation*}
E^\mathbb{Q}\left[e^{-r\delta}\max\{S_ne^{R}-K, 0\}\right] = c(S_n, K, r, \delta, \boldsymbol{\theta}). \label{eq:optionconmod}
\end{equation*}
which will, however, create bias in cases with small sample sizes because of the projected asset price. The difference between the magnitudes of $S_n$, thus the additional constraint, with the two constraints derived from considering the characteristic function of the return series may produce unstable numerical estimates.

\subsection{Model Setup}
Denote $p_1(t), \ldots, p_n(t)$ be probability weights allocated to the residuals $\{\boldsymbol{g}_j(t; \theta)\}_{j=1, \ldots, n}$, where
\begin{equation}
\boldsymbol{g}_j(t;\boldsymbol{\theta}) = \left(\begin{array}{c}
                                          \cos(t R_j) - \phi^R(t;\boldsymbol{\theta})\\
                                          \sin(t R_j) - \phi^I(t;\boldsymbol{\theta})\\   
                                          e^{-r\delta}\max\{e^{R_j}-K, 0\}
                                          \frac{d\mathbb{Q}}{d\mathbb{P}}(R_j;\boldsymbol{\theta})
                                          -c(m, r, \delta, \boldsymbol{\theta})                    
                                          \end{array} \right).
\label{eq:g}
\end{equation}

An empirical likelihood for $\boldsymbol\theta$ at $t$ is given by
\begin{equation}
L_n(\tau, \boldsymbol\theta) = \prod^n_{j=1} p_j(t), \label{eq:emplikelihood}
\end{equation}
subject to constraints $\sum^n_{j=1}p_j(t) = 1$ and $\sum^n_{j=1}p_j(t)\boldsymbol{g}_j(t; \boldsymbol\theta) = \boldsymbol{0}$. Applying Lagrange-multiplier approach as we usually see in maximum empirical likelihood derivation, we see that \eqref{eq:emplikelihood} is maximized when
\begin{equation*}
p_j(t) = \frac{1}{n}\frac{1}{\boldsymbol{\lambda}(t; \boldsymbol{\theta})^\prime \boldsymbol{g}_j(t; \boldsymbol{\theta})},
\end{equation*}
where $\boldsymbol\lambda(t; \boldsymbol\theta)$ is a Lagrange multiplier in $\mathbb{R}^k$ satisfying
\begin{equation*}
\boldsymbol Q_{1n} =: \frac{1}{n}\sum^n_{j=1}\frac{\boldsymbol{g}_j(t; \boldsymbol\theta)}{1+\boldsymbol{\lambda}(t; \boldsymbol{\theta})^\prime\boldsymbol{g}_j(t; \boldsymbol{\theta})} = \boldsymbol 0.
\end{equation*}

Hence, the local log empirical likelihood ratio becomes
\begin{equation*}
\ell_n(t; \boldsymbol\theta) = 2\sum^n_{j=1}\log\left\{1+\boldsymbol\lambda(t; \boldsymbol\theta)^\prime \boldsymbol{g}_j(t; \boldsymbol\theta)\right\}.
\end{equation*}

Like \cite{Chan_etal-2009-JASA}, we consider integrating $\ell_n(t; \boldsymbol\theta)$ against a probability weight $\pi(t)$, an integrated empirical likelihood ratio for $\boldsymbol\theta$ is given by
\begin{equation*}
\ell_n(\boldsymbol\theta) = \int_{t \in \mathbb{R}} \ell_n(t; \boldsymbol\theta)\pi(t)dt.
\end{equation*}

The maximum empirical likelihood estimator (MELE) for $\theta$ is defined as
\begin{equation*}
\widehat{\boldsymbol\theta}_n = \argmin_{\boldsymbol\theta \in \boldsymbol \Theta} \ell_n(\boldsymbol\theta).  
\end{equation*} 
Remark: $\arg\min$ is considered because $-2$ has been multiplied to the EL ratio $\ell_n(\boldsymbol\theta)$.\\
   
The above estimation procedure can be easily extended to situation in which there is more than one option traded in the market. In other words, options with the same maturity but different strikes can be added as additional constraints. For multiple strike constraints, say there are $k$ European calls with moneynesses $m_i~(i=1, \ldots, k)$ respectively, one can simply rewrite \eqref{eq:g} as 

\begin{equation*}
\boldsymbol g_j(t;\boldsymbol\theta) = \left(\begin{array}{c}
                                      \cos(t R_j) - \phi^R(t;\boldsymbol\theta)\\
                                      \sin(t R_j) - \phi^I(t;\boldsymbol\theta)\\   
                                      e^{-r\delta}\max\{e^{R_j}-m_1, 0\}
                                      \frac{d\mathbb{Q}}{d\mathbb{P}}(R_j;\boldsymbol{\theta})
                                      -c(m_1, r, \delta, \boldsymbol{\theta})\\                    
                                      \vdots\\
                                      e^{-r\delta}\max\{e^{R_j}-m_m, 0\}
                                      \frac{d\mathbb{Q}}{d\mathbb{P}}(R_j;\boldsymbol{\theta})
                                      -c(m_m, r, \delta, \boldsymbol{\theta})                   
                                      \end{array} \right)
\end{equation*}
through which we can obtain $\widehat{\boldsymbol\theta}_{EL}$ using the same estimation procedure.
\subsection{Multiple-period Model}
The above framework can be further extended to incorporate options with different strikes as well as different maturities. Similar to the single-period case set-up, suppose we have observed a series of returns $\{R_j\}_{j=1, \ldots, n}$ with the current asset price $S_n$. In addition, we can also obtain prices for calls with different maturities M and moneynesses $m$.\\
   
The procedure will follow closely to the methodology proposed in Section 2.2.  We start from the simplest case in which there are two groups of calls: one group contains $N_1$ calls with different moneynesses but the same maturity $\delta$ while the other group containing $N_2$ calls with different moneynesses but the same maturity $2\delta$.\\
   
The single-period case can be dealt as what we have done in Section 2. For the double-period model, we can view each pair of consecutive (non-overlapping) returns as a single observation. In this case, the double-period model can be reduced to the single-period model with $n/2$ number of observations. Essentially, it means
   \begin{equation*}
   R^{(2)}_j = R_{2j-1}+R_{2j},~~~~~~j = 1, \ldots, n/2
   \end{equation*}
   The corresponding set of estimating equations for the calls that mature in $2\delta$ can be written as 
   \begin{equation*}
   \boldsymbol g^{(2)}_j(t;\boldsymbol\theta) = \left(\begin{array}{c}
                           \cos(t R^{(2)}_j) - \phi^R(t;\boldsymbol\theta)\\
                           \sin(t R^{(2)}_j) - \phi^I(t;\boldsymbol\theta)\\   
                           e^{-r\delta}\max\{e^{R^{(2)}_j}-m^{(2)}_1, 0\}
                           \frac{d\mathbb{Q}}{d\mathbb{P}}(R^{(2)};\boldsymbol{\theta})
                           -c^{(2)}(m^{(2)}_1, r, \delta, \boldsymbol{\theta})\\                    
                           \vdots\\
                           e^{-r\delta}\max\{e^{R^{(2)}_j}-m^{(2)}_{N_2}, 0\}
                           \frac{d\mathbb{Q}}{d\mathbb{P}}(R^{(2)};\boldsymbol{\theta})
                           -c^{(2)}(m^{(2)}_{N_2}, r, \delta, \boldsymbol{\theta})
                           \end{array} \right),
   \end{equation*}           
   where $c^{(2)}$, and $m^{(2)}$ denote respectively the call prices and their corresponding moneynesses.\\
   
   We can define $p^{(2)}_j(t)$, $\boldsymbol Q^{(2)}_{1n}$ and $l^{(2)}_n(t;\boldsymbol\theta)$ accordingly for double-period model. In general, we can also extend above extension to multiple-period case in which $g$ becomes
   \begin{equation*}
   \boldsymbol g^{(k)}_j(t;\boldsymbol\theta) = \left(\begin{array}{c}
                           \cos(t R^{(k)}_j) - \phi^R(t;\boldsymbol\theta)\\
                           \sin(t R^{(k)}_j) - \phi^I(t;\boldsymbol\theta)\\   
                           e^{-r\delta}\max\{e^{R^{(k)}_j}-K^{(k)}_1, 0\}\frac{d\mathbb{Q}}{d\mathbb{P}}(R^{(k)}_j;\boldsymbol{\theta})
                           -c^{(k)}(m^{(k)}_1, r, \delta, \boldsymbol{\theta})\\                    
                           \vdots\\
                           e^{-r\delta}\max\{e^{R^{(k)}_j}-K^{(k)}_{m^{(k)}}, 0\}\frac{d\mathbb{Q}}{d\mathbb{P}}(R^{(k)}_j;\boldsymbol{\theta})
                           -c^{(k)}(m^{(k)}_{m^{(k)}}, r, \delta, \boldsymbol{\theta})
                           \end{array} \right),
   \end{equation*}    
   where $R^{(k)}_j = R_{(j-1)k+1} + \ldots + R_{kj}$.\\
         
    Following the idea of \cite{Chan_etal-2009-JASA}, we try to express our overall likelihood as a sum of all the sub-empirical likelihood. The maximum likelihood estimator for $\boldsymbol\theta$ can be defined similarly as
   \begin{equation}
   \widehat{\boldsymbol\theta}_n = \argmin_{\boldsymbol\theta \in \boldsymbol \Theta} \sum^{n_O}_{j=1} \ell^{(j)}_n(\boldsymbol\theta),
   \end{equation}
   where $n_O$ denotes the number of unique maturities of the options observed. Readers should be noted that, for simplicity, we just use the call prices as constraints. One can use any other option prices as long as they can write down the estimating equations. The inclusion of the puts may not help estimation due to the put-call parity. This methodology, of course, performs worse when the maximum maturity becomes long that leads to a huge reduction of the number of observations. One should note that, however, only options with short maturities are traded actively. These options, meanwhile, provide the most up-to-date, thus useful, information about the parameters.

\section{Models and Examples}
In this section, three commonly used models with known characteristic functions are considered. Discretely observed data are used to investigate the performance of the proposed empirical likelihood estimator to provide an accurate estimate of the unknown parameters of the continuous time models studied. 
\subsection{Black-Scholes Model}
Suppose the stock price $S_t$ follow the geometric Brownian motion

\begin{equation}
d\log S_t = (\mu - \frac{\sigma^2}{2})dt + \sigma dW_t,
\label{eq:st}
\end{equation}

where $W_t$ is a $\mathbb{P}$-Brownian motion. 
Again, we denote the historical returns of the previous $n$ trading period as $R_{j} = \log S_{j} - \log S_{(j-1)}, j = 1, \ldots, n$, we know that for each $j$, 
$$R_j \stackrel{iid}{\sim} \mathcal{N}\left((\mu-\frac{\sigma^2}{2})\delta, \sigma^2\delta\right).$$
The characteristic function of $S_j$ is given by

\begin{equation*}
\phi(t; \boldsymbol\theta) = \exp\{\delta(it(\mu-\sigma^2/2) - \sigma^2t^2/2)\},
\end{equation*}

where $\boldsymbol\theta=(\mu, \sigma)$.
Hence, for any $j = 1, \ldots, n$,

\begin{equation}
E^\mathbb{P}[e^{itR} - \exp\{\delta(it(\mu-\sigma^2/2) - \sigma^2t^2/2)\} ] = 0
\label{eq:char}
\end{equation}

is an estimating equation for $\boldsymbol\theta$.\\
  
  In addition to the return series, we also observe option prices traded at time $n\delta$, each of them expires in the next period of length $\delta$: $\{c(m_j, r, \boldsymbol\theta)\}_{j = 1, \ldots, k}$. From these $k$ option prices, we can write down an estimating equation for the parameters $\boldsymbol\theta=(\mu, \sigma)$: 
  \begin{eqnarray*}
  0 & = &E^\mathbb{P}\left[e^{-r\delta}\max\{e^{R}-m, 0\}\frac{d\mathbb{Q}}{d\mathbb{P}}(R;\boldsymbol{\theta})-c(m, r, \delta, \boldsymbol{\theta})\right],
  \end{eqnarray*}
   where, if \eqref{eq:st} holds,
   \begin{eqnarray*}
   \frac{d\mathbb{Q}}{d\mathbb{P}}(R; \boldsymbol\theta)
   & = & \left[\frac{1}{\sqrt{2\pi\sigma^2\delta}}\exp\left\{-\frac{(R-(r-\sigma^2/2)\delta)^2}{2\sigma^2\delta}\right\}\right]
   \left[\frac{1}{\sqrt{2\pi\sigma^2\delta}}\exp\left\{-\frac{(R-(\mu-\sigma^2/2)\delta)^2}{2\sigma^2\delta}\right\}\right]^{-1}\\
   & = & \exp\left\{-\frac{1}{2\sigma^2\delta}\left[(R-(r-\sigma^2/2)\delta)^2-(R_{j}-(\mu-\sigma^2/2)\delta)^2\right]\right\}\\
   & = & \exp\left\{\frac{r-\mu}{\sigma^2}R - \frac{r^2-\mu^2}{2\sigma^2}\delta+\frac{(r-\mu)\delta}{2}\right\}.
   \end{eqnarray*}
   This leads to the following estimating equation
   \begin{equation}
   E\left[\left\{e^{-r\delta}\max\{e^R-m,0\} - c(m, r, \delta, \boldsymbol\theta)\right\}\exp\left\{\frac{r-\mu}{\sigma^2}R - \frac{r^2-\mu^2}{2\sigma^2}\delta+\frac{(r-\mu)\delta}{2}\right\}\right] = 0.
   \label{eq:optionsBS}
   \end{equation}  
   \subsection{Black-Scholes Model with Merton Jumps (BS-MJ)}
   Empirical studies suggest that log return sequences usually exhibit skewness and an excess kurtosis (compared with a normal distribution). In order to devise a model that can provide a better fit to the financial market data, \cite{Merton-1976-JoFE}, believing that the Black-Scholes solution is not valid as the stock prices dynamics should not be presented by a stochastic process with a continuous path, proposed Black-Scholes Model with jumps (BS-MJ), which is specified as follows:

   \begin{equation}
   dS_t = (\mu - \lambda\kappa)S_t dt + \sigma S_t dW_t + (J_t - 1)S_t dN_t,
   \label{eq:StBSMJ}
   \end{equation}

   where $N_t$ is a Poisson process with intensity parameter $\lambda > 0$ and $J_t$ is the jump size following a lognormal distribution $\log-\mathcal{N}(\mu_J, \sigma^2_J)$ and is independent of $W_t$. $\lambda\kappa := \lambda E[J_t - 1] = \lambda(\exp\{\mu_J + \sigma^2_J/2\} - 1)$ is the compensator of the compound Poisson process $(J_t - 1)S_t dN_t$.\\
   
By Ito's lemma for jump diffusion processes (see \cite{Shreve-2004-Springer}), \eqref{eq:StBSMJ} can be rewritten as
   
  \begin{equation}
   d\log S_t = (\mu - \lambda\kappa - \sigma^2/2)dt + \sigma dW_t + \log J_t dN_t
   \label{eq:logStBSMJ},
   \end{equation}   
under the physical measure $\mathbb{P}$. Despite the fact that there is no closed form density for $\log S_t$, its characteristic function is given as follows:
   
   \begin{equation}
   \phi(t; \boldsymbol{\theta}) = \exp\left\{\delta\left[it(\mu-\lambda\kappa-\sigma^2/2)-\sigma^2t^2/2+\lambda\left(e^{i\mu_Jt-\sigma_J^2 t^2/2}-1\right)\right]\right\}.
   \label{eq:cfBSMJ}
   \end{equation}
By constructing a hedging portfolio, \cite{Merton-1976-JoFE} proposed that the European call option price on an equity that follows the dynamics given by \eqref{eq:logStBSMJ} $V(S_t, t)$ should be the solution of 
   \begin{equation*}
   \frac{\partial V}{\partial t} + \frac{\sigma^2 S_t^2}{2}\frac{\partial^2 V}{\partial S_t^2} + rS_t\frac{\partial V}{\partial S_t} - rV + \lambda E[V(J_tS_t, t) - V(S_t, t)] 
   -\lambda S_t \frac{\partial V}{\partial S_t} E[J_t - 1] = 0,
   \end{equation*}
   which is equal to
   \begin{eqnarray*}
   V(t, S_t) & = & e^{-r\tau}E^{\mathbb{Q}_M}[\max\{S_(t+\delta)-K,0\}|S_t]\\
             & = & \sum_{n\geq0} \frac{e^{-\bar{\lambda}\delta}(\bar{\lambda}\delta)^n}{n!}V^{BS}\left(\delta, S_t; \sigma_n, r_n\right),
   \end{eqnarray*}
   with
   \begin{eqnarray*}
   \bar{\lambda} & = & \lambda(1+\kappa) = \lambda\exp\{\mu_J + \sigma_J^2/2\}\\
   \sigma_n & = & \sqrt{\sigma^2 + n\sigma_J^2/\delta}\\
   r_n & = & r - \lambda\kappa+\frac{n\mu_J+n\sigma_J^2/2}{\delta}\\
   V^{BS}(\delta, S, \sigma, r) & = & S\mathcal{N}\left(\frac{\log(S/K)+(r+\sigma^2/2)\delta}{\sigma\sqrt{\delta}}\right)-Ke^{-r\delta}\mathcal{N}\left(\frac{\log(S/K)+(r-\sigma^2/2)\delta}{\sigma\sqrt{\delta}}\right).
   \end{eqnarray*}
   
Again, we need to compute the Radon-Nikodym derivative between the two measures $\mathbb{P}$ and $\mathbb{Q}_M$. Using the inverse Fourier transform formula, we can express the density of the Merton's jump diffusion model under physical measure $\mathbb{P}$ as follows:
   
   \begin{eqnarray*}
   f(x) & = & \frac{1}{2\pi}\int^\infty_{-\infty}\phi(t; \mathbf{\theta})e^{-itx}dt\\
        & = & \frac{e^{\lambda\delta}}{2\pi}\int^\infty_{-\infty}e^{i\delta t(\mu-\lambda\kappa-\sigma^2/2)-\sigma^2t^2/2}
              \exp\{e^{it\mu_J - t^2\sigma_J^2/2}\lambda\delta\}dt\\
        & = & \frac{e^{\lambda\delta}}{2\pi}\int^\infty_{-\infty}e^{i\delta t(\mu-\lambda\kappa-\sigma^2/2)-\sigma^2t^2/2}
              \sum_{n\geq 0}\frac{(\lambda\delta)^n}{n!}(e^{it\mu_J-t^2\sigma_J^2/2})^ndt\\
        & = & \frac{e^{\lambda\delta}}{2\pi}\sum_{n\geq 0}\frac{(\lambda\delta)^n}{n!}
              \int^\infty_{-\infty}\exp\{i\delta t(\mu-\lambda\kappa-\sigma^2/2)-\sigma^2t^2/2-itx+int\mu_J-nt^2\sigma_J^2/2\}dt.
   \end{eqnarray*}
   
Using the identity that $\int^\infty_{-\infty}e^{-az^2+ibz}dz = \sqrt{\frac{\pi}{a}}e^{-b^2/4a}$, putting $a = \delta\sigma^2/2 + n\sigma_J^2/2$, $b = (\mu-\sigma^2/2 - \lambda\kappa)\delta - x + n\mu_J$, we can write
   \begin{equation*}
   f(x) = \frac{e^{-\lambda\delta}}{\sqrt{2\pi}}\sum_{n\geq 0}\frac{(\lambda\delta)^n}{n!}
          \frac{e^{-\frac{[(\mu-\sigma^2/2-\lambda\kappa)\delta+n\mu_J-x]^2}{2(\delta\sigma^2+n\sigma_J^2)}}}{\sqrt{\delta\sigma^2+n\sigma_J^2}},
   \end{equation*}  
which is a fast converging sequence. So, the Radon-Nikodym derivative required is

   \begin{equation*}
   \frac{d\mathbb{Q}_M}{d\mathbb{P}}(R_j; \mathbf{\theta})
       = \left[\sum_{n\geq 0}\frac{(\lambda\delta)^n}{n!}\frac{e^{-\frac{[(\mu-\sigma^2/2-\lambda\kappa)\delta+n\mu_J-R_j]^2}
         {2(\delta\sigma^2+n\sigma_J^2)}}}{\sqrt{\delta\sigma^2+n\sigma_J^2}}\right]
         \left[\sum_{n\geq 0}\frac{(\lambda\delta)^n}{n!}\frac{e^{-\frac{[(r-\sigma^2/2-\lambda\kappa)\delta+n\mu_J-R_j]^2}
         {2(\delta\sigma^2+n\sigma_J^2)}}}{\sqrt{\delta\sigma^2+n\sigma_J^2}}\right]^{-1}.
   \end{equation*}

   In other words, the corresponding estimating equation that is derived from an option is given by

   \begin{eqnarray*}
   0 & = & E\left[\left(e{^{-r\delta}\max\{e^{R}}-m,0\}- \sum_{n\geq0} \frac{e^{-\bar{\lambda}\delta}(\bar{\lambda}\delta)^n}{n!}V^{BS}\left(\delta, S_t; \sigma_n, r_n\right)\right)\right.\nonumber\\
     &   & \left.\left(\sum_{n\geq 0}\frac{(\lambda\delta)^n}{n!}\frac{e^{-\frac{[(\mu-\sigma^2/2-\lambda\kappa)\delta+n\mu_J-R_j]^2}
         {2(\delta\sigma^2+n\sigma_J^2)}}}{\sqrt{\delta\sigma^2+n\sigma_J^2}}\right)
         \left(\sum_{n\geq 0}\frac{(\lambda\delta)^n}{n!}\frac{e^{-\frac{[(r-\sigma^2/2-\lambda\kappa)\delta+n\mu_J-R_j]^2}
         {2(\delta\sigma^2+n\sigma_J^2)}}}{\sqrt{\delta\sigma^2+n\sigma_J^2}}\right)^{-1}\right].
   \label{eq:optionsBSMJ}
   \end{eqnarray*}   
   To generate $\log S_t$ from \eqref{eq:logStBSMJ}, we use a sequence of Bernoulli processes to approximate the Poisson jump process. Discretised sample paths can be generated through

   \begin{equation*}
   \log S_{n+1} = \log S_n + (\mu-\lambda\kappa-\sigma^2/2)\delta + \sigma\sqrt{\delta}Z + \sum^{200}_{l=1}N_lJ_l,   
   \end{equation*}
where $Z$ denotes a standard normal random variable, $J_l \sim \mathcal{N}(\mu_J, \sigma_J^2)$ and\\
 $N_l \sim \text{Bernoulli}\left((\lambda\delta/200)\exp\{-\lambda\delta/200\}\right)$.
   \subsection{Double-Exponential Jump Model}
      \cite{Kou-2002-MS} proposed a jump-diffusion similar Merton's, where the jump size is double-exponentially distributed. The double-exponential jump diffusion (DEJD) model is designed to capture the leptokurtic feature of the empricial return distributions as well as the volatility smile in option markets which cannot be successfully modeled by BS-MJ model. The canonical decomposition of the driving process of Kou's model is
   \begin{equation}
   dS_t = \mu S_t dt + \sigma S_t dW_t + S_td\left(\sum^{N(t)}_{i=1}(V_i -1)\right),
   \label{eq:kou}
   \end{equation}
   where $W_t$ is a standard Brownian motion, $N(t)$ is a Poisson process with rate $\lambda$ and $\{V_i\}$ is a sequence of independent identically distributed non-negative random variables such that $Y \triangleq \log(V)$ has an asymmetric double exponential distribution with the density
   \begin{equation*}
   f_Y(y) = p\eta_1e^{-\eta_1y}\mathbf{1}_{\{y \geq 0\}} + (1-p)\eta_2e^{\eta_2 y}\mathbf{1}_{\{y<0\}},
   \end{equation*}
   with $\eta_1, \eta_2 > 0$, 
   where $p \geq 0$ represent the probabilities of upward and downward jumps, i.e. 
   \begin{equation*}
   \log(V) = Y =_d \begin{cases} \xi^+~~&,\text{with probability}~~$p$\\ -\xi^-~~&,\text{with probability}~~$1-p$,\end{cases}
   \end{equation*}
   where $\xi^+$ and $\xi^-$ are exponential random variables with means $\eta^{-1}_1$ and $\eta^{-1}_2$ respectively. In model \eqref{eq:kou}, all sources of randomness, $N(t)$, $W(t)$ and $Y$'s are assumed to be independent.\\
   
  The analytical solution of a call option whose price is deteremined by an underlying asset that is driven by DEJD model also incorporates a psychological interpretation of investors. As we can see in \eqref{eq:kou}, this model has six parameters, namely $\mu$, the drift parameter, $\sigma$, the diffusion volatility, $\lambda$, the Poisson rate, $p$, the probability of having an upward jump, $\eta_1$, the rate of an upward exponential jump and $\eta_2$, the rate of a downward exponential jump. By incorporating option prices observed with different strikes and maturities, we can improve the estimation, compared with incorporating merely the characteristic function of the model. In addition, the option prices used can also enable the estimation of the parameters involved in the utility function.\\

   Given \eqref{eq:kou}, one can write down the dynamic of $d\log S_t$ by using Ito's Lemma:
   \begin{equation*}
   d\log S_t = \left(\mu - \frac{\sigma^2}{2}\right)dt + \sigma dW_t + d\left(\sum^{N(t)}_{i=1}Y_i\right),
   \end{equation*}
   from which we can derive the characteristic function of $\log S_t$ (see \cite{Cont_Tankov-2004-CRC}) under the risk-neutral probability measure without taking the jump risk into account:
   \begin{equation*}
   \phi_T(u) = E[e^{iu\log S_t}] = \exp\left\{t\left(\left[\log S_0 + r -\frac{\sigma^2}{2}\right]u + iu\lambda\left[\frac{p}{\lambda_+-iu}-\frac{1-p}{\lambda_-+iu}\right]\right)\right\},
   \end{equation*}
   since the L\'evy density of the jump is
   \begin{equation*}
   \nu(x) = p\lambda\eta_1e^{-\eta_1x}\mathbf{1}_{\{x>0\}} + (1-p)\lambda\eta_2e^{\eta_2 x}\mathbf{1}_{\{x\leq0\}}.
   \end{equation*}
   The corresponding European call price can be obtained via \cite{Carr_Madan-1999-JoCF} method, which is specified as follows:
   \begin{equation*}
   C(S_0, K, T, r) = \frac{e^{-\alpha \log K}}{2\pi}\int^\infty_{-\infty} e^{-iu\log K}\frac{e^{-rT}\phi_T(u-(\alpha+1)i)}{\alpha^2+\alpha - u^2 + i(2\alpha+1)u}du.
   \end{equation*}
   Using the independence between the expo- nential and normal distributions used in the model and formulae for the sum of double exponential random variables, \cite {Kou-2002-MS} obtains the probability density function of the return, which can be approximated by the following density function:
   \begin{eqnarray*}
   f_R(x; \mu) & := & \frac{1-\lambda\delta}{\sigma\sqrt{\delta}}\phi\left(\frac{x-\mu\delta}{\sigma\sqrt{\delta}}\right)\\
        & = & \lambda\delta\left\{p\eta_1e^{(\sigma^2\eta^2_1\delta)/2}e^{-(x-\mu\delta)\eta_1}\Phi\left(\frac{x-\mu\delta - \sigma^2\eta_1\delta}{\sigma\sqrt{\delta}}\right)\right.\\
        &  & + q\eta_2e^{(\sigma^2\eta_2^2\delta)/2}e^{(x-\mu\delta)\eta_2} \left.\Phi\left(-\frac{x-\mu\delta+\sigma^2\eta_2\delta}{\sigma\sqrt{\delta}}\right)\right\},
   \end{eqnarray*}
   which can be used to define the Radon-Nikodym derivative to adjust for the difference between a risk-free probability measure (in Merton's sense) and the physical measure since
   \begin{equation*}
   \frac{d\mathbb{Q}^M}{d\mathbb{P}} = \frac{f_R(x;r)}{f_R(x;\mu)}.
   \end{equation*}
   The estimating equaiton derived from the option price is 
   \begin{eqnarray*}
   0 & = & E^{\mathbb{P}}\left[\left(e{^{-r\delta}\max\{e^{R}}-m,0\}- C(S_n, K, T, r)/S_n\right)\frac{d\mathbb{Q}^M}{d\mathbb{P}} (R; \boldsymbol\theta)\right].\nonumber
   \end{eqnarray*}
   \subsection{With Jump Risk Premium}
   \cite{Kou-2002-MS} considered a typical rational expectations economy (\cite{Lucas-1978-Econometrica}) in which a representative investor has the utility function of the special form, as in \cite{Naik_Lee-1990-RFS}:
   \begin{equation}
   U(c,t) = \begin{cases}
            e^{-\kappa t}c^\alpha/\alpha &, \text{ if } 0 < \alpha < 1 \\
            e^{-\kappa t}\log(c) &, \text{ if } \alpha = 0, \label{eq:utility}
            \end{cases}
   \end{equation}
   with $U_c(c,t) \triangleq \frac{\partial U(c,t)}{\partial c}$. The goal of the representative investor is to obtain $\max_cE[\int^\infty_0 U(c(t),t)dt]$. In his model, Kou also assumed $E_t$, an endowment process, which is, under the physical measure $\mathbb{P}$, specified as follows:
   \begin{equation}
   \frac{dE_t}{E_t} = \mu_1 dt + \sigma_1 dW^{(1)}_t + d\left[\sum^{N(t)}_{l=1}(\tilde{V}_l-1)\right]; \label{eq:endow}
   \end{equation}
   given the endowment process \eqref{eq:endow}, the asset price will have the dynamic of the form
   \begin{equation}
   \frac{dS_t}{S_t} = \mu dt + \sigma\{\rho dW^{(1)}_t + \sqrt{1-\rho^2}dW^{(2)}_t\}+d\left[\sum^{N(t)}_{l=1}(V_l -1)\right], \label{eq:koust}
   \end{equation}
   where $dW^{(2)}_t$ is a Brownian motion independent of $dW^{(1)}_t$ and $V_l = \tilde{V}_l^\beta$. Furthermore, $\alpha$ and $\kappa$ in \eqref{eq:utility} are related as follows:
   \begin{equation*}
   \mu = \kappa + (1-\alpha)\left\{\mu_1 - \frac{1}{2}\sigma^2_1(2-\alpha)+\sigma_1\sigma\rho\right\}-\lambda\zeta_1^{(\alpha+\beta-1)},
   \end{equation*}
   where $\zeta_1^{(a)} \triangleq E[\tilde{V}^a - 1]$.\\
   
   It can been shown (see, for example, \cite{Stokey_Lucas-1989-HUP}) that, under mild conditions, the rational expectations equilibrium price, or the ``shadow'' price, of the security $p(t)$, must satisfy the Euler equation
   \begin{equation}
   p(t) = \frac{E[e^{-\theta T(\delta(T))^{\alpha-1}}p(T)|\mathcal{F}_t]}{e^{-\theta t}(\delta(t))^{\alpha-1}}, \quad \forall T \in[t, T_0],
   \end{equation}
   where $U_c$ is the partial derivative of $U$ with respect to $c$.
   To simplify the model, we assume $E_t = S_t$, i.e. $\mu_1 = \mu$, $\sigma_1 = \sigma$ and $\rho = \beta=1$. It follows that, as shown in (10) of \cite{Kou-2002-MS}, the Radon-Nikodym derivative between the risk-free measure $\mathbb{Q}$ and the physical measure $\mathbb{P}$ is given by
   \begin{eqnarray*}
   \frac{d\mathbb{Q}}{d\mathbb{P}}(R_{(j+1)\delta}; \theta) & = & \frac{e^{r(j+1)\delta}U_c(S_{(j+1)\delta},(j+1)\delta))}{e^{rj\delta}U_c(S_{j\delta},j\delta)} = e^{(r-\kappa)\delta}\left[\frac{S_{(j+1)\delta}}{S_{j\delta}}\right]^{\alpha-1}\\
   & = & e^{\delta\left[(1-\alpha)\mu-\frac{1}{2}\sigma^2(1-\alpha)(2-\alpha))\right]}e^{R_{(j+1)\delta}(\alpha-1)}.
   \end{eqnarray*}
   Here $\lambda$ is a Poisson process with rate $\lambda$. The jump sizes $\{Y_1, Y_2, \ldots\}$ are independent identically distributed random variables such that $Y_i = \log(V_i)$. The moment generating function of $X(t):=\log(S_t/S_0)$ can be obtained as 
   \begin{equation*}
   E\left[e^{\theta X(t)}\right] = \exp\left\{G(\theta)t\right\},
   \end{equation*}   
   where $G(x) = \widetilde{\mu} x + \frac{1}{2}x^2\sigma^2 + \lambda\left(E\left[e^{xY}\right]-1\right)$. In the case of Merton's normal jump-diffusion model, 
   \begin{equation*}
   G(x) = \widetilde{\mu} x + \frac{1}{2}x^2\sigma^2 + \lambda\left\{\mu_Jx + \frac{x^2\sigma^2_J}{2}-1\right\};
   \end{equation*}
   and in the case of double exponential jump-diffusion model
   \begin{equation*}
   G(x) = \widetilde{\mu} x + \frac{1}{2}x^2\sigma^2 + \lambda\left(\frac{p\eta_1}{\eta_1 - x}+\frac{(1-p)\eta_2}{\eta_2 + x}-1\right).
   \end{equation*}
   Under the risk-neutral probability $\mathbb{Q}$, we have
   \begin{equation*}
   \widetilde{\mu} = r - \frac{1}{2}\sigma^2 - \lambda\zeta,
   \end{equation*}
   where $\zeta := E\left[e^Y\right] -1$. In the Merton's model 
   \begin{equation*}
   \zeta = E^{\mathbb{Q}}\left[e^Y\right] -1 = \mu_J +\frac{\sigma^2_J}{2} - 1,
   \end{equation*}
   while in the double exponential jump-diffusion model 
   \begin{equation*}
   \zeta = \frac{p\eta_1}{\eta_1-1}+\frac{(1-p)\eta_2}{\eta_2+1}-1.
   \end{equation*}
   \cite{Kou_etal-2005-KER} adpated the method in \cite{Carr_Madan-1999-JoCF}, which is based on a change of the order of integration, to price European call and pution options via Laplace transforms. The Laplace transform with respect to $k$ of $C(S, e^k, r, T)$ is given by
   \begin{eqnarray*}
   \widehat{f}_C(\xi) & := & \int^\infty_{-\infty} e^{-\xi k}C(S, e^k, r, T)dk\\ 
                      & = & e^{-rT}\frac{S^{\xi+1}}{\xi(\xi+1)}\exp\left\{G(\xi+1)T\right\}, \quad \xi >0.
   \end{eqnarray*}
   This leads to the following estimating equation:
   \begin{equation*}
   E\left[\left(e^{-r\delta}\max\left\{e^R - m, 0\right\} - c(m, r, \delta, \boldsymbol\theta)\right)e^{\delta\left[(1-\alpha)\mu-\frac{1}{2}\sigma^2(1-\alpha)(2-\alpha))\right]}e^{R(\alpha-1)}\right] = 0.
   \end{equation*}
   
   Note that the option price $c$ under the double exponential jump diffusion dynamics can also be obtained directly using the method proposed by \cite{Kou-2002-MS},
   \begin{eqnarray*}
   c(m; r, \delta, \boldsymbol\theta) & = & \Upsilon\left(r+\frac{1}{2}\sigma^2-\lambda\zeta, \sigma, \tilde{\lambda}, \tilde{p}, \tilde{\eta}_1, \tilde{\eta}_2; \log(1/m), \delta\right)\\
   & + & me^{-r\delta}\Upsilon\left(r-\frac{1}{2}\sigma^2-\lambda\zeta, \sigma, \tilde{\lambda}, \tilde{p}, \tilde{\eta}_1, \tilde{\eta}_2; \log(1/m), \delta\right),
   \end{eqnarray*}
where the definitions of $\tilde{\lambda}, \tilde{p}, \tilde{\eta}_1, \tilde{\eta}_2$ and $\Upsilon(\cdot)$ can be found in \cite{Kou-2002-MS}.\\

   The characteristic function of a return drived from the price driven by the process \eqref{eq:koust} can be obtained similarly as in Sections 3.2 and 3.3.
   The corresponding estimating equations are thus
   \begin{equation*}
   E[\cos(tR)] - \text{Re}(\phi(t;\boldsymbol\theta)) = E[\sin(tR)] - \text{Im}(\phi(t;\boldsymbol\theta)) = 0.
   \end{equation*}

   \section{Asymptotic Results}
Regularity conditions:
\begin{enumerate}
\item $E[\boldsymbol{g}(t, X_1; \boldsymbol{\theta}_0)\boldsymbol{g}(t, X_1; \boldsymbol{\theta}_0)^\prime]$ is positive definite for $t \in [-a, a], a>0$;
\item $\frac{\partial}{\partial\boldsymbol{\theta}}\boldsymbol{g}(t, x; \boldsymbol{\theta})$ is continuous in a neighbourhood of $\boldsymbol{\theta}_0$, for $t\in[-a, a], x\in\mathbb{R}$;
\item $\sup_{\boldsymbol{\theta}\in\Theta} \|\frac{\partial}{\partial\boldsymbol{\theta}}\boldsymbol{g}(t, x; \boldsymbol{\theta})\| \leq H(t, x)$, where
$$\int^a_{t=-a}\int^\infty_{x=-\infty} H(t,x)dF(x)dG_1(t) < \infty;$$\label{condthree}
\item The rank of $E[\frac{\partial}{\partial\boldsymbol\theta}\boldsymbol g(t, X_1; \boldsymbol\theta_0)]$ is $\min\{2,d\}$ for all $t\in[-a,a]$, where $d$ is the dimension of $\boldsymbol{\theta}$;
\item $\frac{\partial^2}{\partial\boldsymbol{\theta}\partial\boldsymbol{\theta}^\prime}\boldsymbol{g}(t, x; \boldsymbol{\theta})$ is continuous in $\boldsymbol{\theta}$ for $\boldsymbol{\theta}\in\Theta$, $t\in[-a,a]$ and $x\in\mathbb{R}$;
\item $\sup_{\boldsymbol{\theta}\in\Theta}\|\frac{\partial^2}{\partial\boldsymbol{\theta}\partial\boldsymbol{\theta}^\prime}\boldsymbol{g}(t, x; \boldsymbol{\theta})\|\leq H(t,x)$, where $H$ is given in \ref{condthree}.
\end{enumerate}
\begin{prop}
Under conditions 1-4, with probability one, denote $\widehat{\boldsymbol{\theta}} = \arg\min_{\boldsymbol\theta} T_1(\boldsymbol{\theta})$ which satisfies $\|\widehat{\boldsymbol{\theta}}-\boldsymbol{\theta}_0\| \leq n^{-1/3}$,
\begin{eqnarray*}
\boldsymbol{Q}_{1n}(t, \widehat{\boldsymbol{\theta}}, \boldsymbol{\lambda}_1(t;\widehat{\boldsymbol{\theta}})) & = &  0\\
\int^a_{-a}\boldsymbol{Q}_{2n}(t, \widehat{\boldsymbol{\theta}}, \boldsymbol{\lambda}_1(t;\widehat{\boldsymbol{\theta}})) dG_1(t)& = & 0,
\end{eqnarray*}
where 
\begin{eqnarray*}
\boldsymbol Q_{1n}(t; \boldsymbol\theta, \boldsymbol\lambda) & = & \frac{1}{n}\sum^n_{j=1}\frac{\boldsymbol g(t; x_j, \boldsymbol\theta)}{1+\boldsymbol\lambda^\prime\boldsymbol g(t; x_j, \boldsymbol\theta)},\\
\boldsymbol Q_{2n}(t; \boldsymbol\theta, \boldsymbol\lambda) & = & \frac{1}{n}\sum^n_{j=1}\frac{1}{1+\boldsymbol\lambda^\prime\boldsymbol g(t; x_j; \boldsymbol\theta)}\frac{\partial \boldsymbol g^\prime(t, x_j; \boldsymbol\theta)}{\partial\boldsymbol\theta}\boldsymbol\lambda.
\end{eqnarray*}
\end{prop}
\begin{prop}
Under conditions 1-6, for the estimator $\widehat{\boldsymbol{\theta}}$ given in Proposition 2.1, we have as $n\rightarrow\infty$, 
\begin{eqnarray*}
\sqrt{n}(\widehat{\boldsymbol{\theta}}-\boldsymbol{\theta}_0) & = & -\left\{\int^a_{-a}\boldsymbol{s}_{21}(t)\boldsymbol{s}^{-1}_{11}(t)\boldsymbol{s}_{12}(t)dG_1(t)\right\}^{-1}\\
&& \times \left\{\int^a_{-a}\boldsymbol{s}_{21}(t)\boldsymbol{s}_{11}^{-1}(t)\sqrt{n}\boldsymbol{Q}_{1n}(t;\boldsymbol{\theta}_0, 0)dG_1(t)\right\}+o_p(1)\\
& \rightarrow_d & N(\boldsymbol{0}, \boldsymbol{\Sigma}),
\end{eqnarray*}
where 
\begin{eqnarray*}
\boldsymbol{s}_{11}(t) & = & -E[\boldsymbol{g}(t, R_1; \boldsymbol{\theta}_0)\boldsymbol{g}(t, R_1; \boldsymbol{\theta}_0)^\prime],\\
\boldsymbol{s}_{12}(t) & = & \boldsymbol{s}_{21}^\prime(t) = E[\frac{\partial}{\partial \boldsymbol{\theta}}g(t, R_1; \boldsymbol{\theta}_0)],\\
\boldsymbol{\Sigma} & = & \left\{\int^a_{-a}\boldsymbol{s}_{21}(t)\boldsymbol{s}^{-1}_{11}(t)\boldsymbol{s}_{12}(t)dG_1(t)\right\}^{-1}\\
&& \times \left\{\int^a_{-a}\int^a_{-a}\boldsymbol{s}_{21}(t_1)\boldsymbol{s}^{-1}_{11}(t_1)\boldsymbol{\Gamma}(t_1, t_2)\boldsymbol{s}^{-1}_{11}(t_2)\boldsymbol{s}_{12}(t_2)dG_1(t_1)dG_2(t_2)\right\}\\
&& \times \left\{\int^a_{-a}\boldsymbol{s}_{21}(t)\boldsymbol{s}^{-1}_{11}(t)\boldsymbol{s}_{12}(t)dG_1(t)\right\}^{-1},
\end{eqnarray*}
where
$$
\boldsymbol{\Gamma}(t_1, t_2) = E[\boldsymbol{g}(t_1, R_t; \boldsymbol{\theta}_0)\boldsymbol{g}(t_1, R_t; \boldsymbol{\theta}_0)^\prime].
$$
\end{prop}
\begin{corollary}
\label{thm:col1}
When $k_1>k_2$, the asymptotic variance $\boldsymbol\Sigma = \boldsymbol\Sigma_k$ of $\sqrt{n}(\widehat{\boldsymbol\theta}-\boldsymbol\theta)$ cannot decrease if an estimating equation is dropped.
\end{corollary}
    
   \section{Numerical Results}
   \subsection{Simulations}
   For each model, 500 sample paths with size $n = 125, 250, 500$ and $1000$ starting at initial value $\log S_0 = 100$ with frequency $\delta = 1/52$ were simulated. Similar to Chan et al. (2009) approach, we also choose the uniform weight function $G(t)$ and $l_n$ can be approximated by the Riemann sum of $l_n(t)$ evaluated at $t \in [-5.0, 5.0]$ with the number of grids set to be 100. \footnote{In Chan et al. (2009), they chose the interval to be $[-0.5, 0.5]$. In the simulation studies, we found that using $[-0.5, 0.5]$ produced poor estimations. A wider interval chosen allows the data to provide more information about the parameter values.} Simulation results for BS, BS-MJ and DEJD are tabulated in Tables \ref{table:simBS},\ref{table:simBSMJ} and \ref{table:simDEJD} respectively. As we can see from the simulation results, by incorporating more option prices, the estimated standard deviation of the estimates are reduced, which is due to the result of Corollary \ref{thm:col1}.\\

   
   
   \subsection{Case Study}
   We examine empirically whether the proposed methodology can be applied to the real data set and what insights call prices can reveal when we incorporate them into the model. Historical S\&P 500 index values and corresponding call option prices between 2 Janurary 1987 and 31 December 2008 were downloaded from Wharton Research Data Services (WRDS). The index is sampled at 4-day frequency and in total we have 1,260 data points. The duration of four days is chosen so as to match the time to maturity of the call prices. The calls were traded on Chicago Board Options Exchange (CBOE).\\

We included, in our simulations, from one to four call prices that were most frequently traded on the last day of our analyses so as to reflect the market information on that particular trading day. The mean annual rate of return is 0.0531 with the associated volatility equals 0.1328. In addition to the market crash of 1987, the tech- and credit-bubble between the late 90's and mid 2000's as well as September 11 attack in 2001, the sample period also covered the recent Lehmann Brother's collapse as a result of credit crunch in 2008. In particular, our data analysis was done with the last day selected as September 29, 2008 - the day on which the largest single day plunge was recorded shortly after Lehman brothers' and Washington Mutual's bankrupcy. Furthermore, on that day, the Volatility S\&P (VIX), a measure of market volatility, has the record highest jump in history. Estimated values of the parameters and the associated esimated asymptotic variances are tabulated in Tables 6 - 8.\\

It can be seen from the tables that by incorporaing constraints due to observed option prices, one can lower the variance of the estimates. It should be also noted that in order for \cite{Chan_etal-2009-JASA}'s approach to achieve the same magnitude of variance as what we can see by including additionally one call price, the sample size should have to be roughly doubled. In other words, using call prices as constraints reduces the required sample size at the expense that the equity price dynamics are specified by a particular model. Since the option prices are considered as a summary of the current market view on the underlying equity price dynamics, our methodology can successfully capture more updated estimate of the current market condition. This can be seen in the data analysis results in which the volatility and/or jump size estimates are both larger than the estimates that \cite{Chan_etal-2009-JASA} provided, which can be interpreted as the consequence because of the late-2000's financial crisis. Finally, we comment that, due to the small number of option prices included, it is challenging to produce an accurate estimate $\alpha$, the risk-preference parameter of investors of which informaiton can be only derived from the option prices. 

   \section{Conclusion}
   L\'evy processes are an excellent tool for modeling price processes in mathematical finance. Its popularity arises from its flexibility and simple structure in comparison with general semimartingales. Estimation for L\'ey processes are challenging statistical inference problems because of the lack of analytical expression for the transitional density function. Inspired by \cite{Chan_etal-2009-JASA} that uses integrated empirical likelihood approach for parameter estimation, we propose in this paper incorporating call prices as constraints in addition to using the characteristic function associated with the process. This method provides a more efficient estimate that can reflect the recent market condition more accurately which is demonstrated via simulations and real data analyses. The idea of using derivative prices as one of the estimating equations is not restricted to call prices only; in fact, any price that can be expressed in terms of expectation of an independent random variable that follows the same distribution as specified by the underlying process are eligible for being included as one of constraints. The approach, therefore, has robust theoretical and versatility for a wide range of processes including processes with jump components.

   \section{Appendix}
   \begin{proof}[Proof of Proposition 4.1]\quad
   It follows closely the proof of lemma 1 of \cite{Qin_Lawless-1994-AoS}.
   \end{proof}
   \begin{proof}[Proof of Proposition 4.2]\quad
   Similar to \cite{Chan_etal-2009-JASA}, we can show that 
   \begin{eqnarray*}
   \frac{\partial}{\partial \boldsymbol\theta} \boldsymbol Q_{1n}(t; \boldsymbol\theta; 0) & \stackrel{P}{\rightarrow} & s_{12}(t),\\
   \frac{\partial}{\partial \boldsymbol\lambda^\prime} \boldsymbol Q_{1n}(t; \boldsymbol\theta; 0) & \stackrel{P}{\rightarrow} & s_{11}(t),\\
   \frac{\partial}{\partial \boldsymbol\theta} \boldsymbol Q_{2n}(t; \boldsymbol\theta; 0) & = & \boldsymbol 0,\\
   \frac{\partial}{\partial \boldsymbol\lambda^\prime} \boldsymbol Q_{2n}(t; \boldsymbol\theta; 0) & \stackrel{P}{\rightarrow} & s_{21}(t)
   \end{eqnarray*}
   uniformly in $t\in[-a,a]$. Denote $\|\widehat{\boldsymbol\theta}-\boldsymbol\theta_0\| + \sup_{t\in[-a,a]}\|\boldsymbol\lambda_1(t; \widehat{\boldsymbol\theta})\|$. Then, we can expand $\boldsymbol Q_{1n}$ and $\boldsymbol Q_{2n}$ using Taylor series expansions and yield
   \begin{eqnarray*}
   \boldsymbol\lambda_1(t; \widehat{\boldsymbol\theta})
   & = & -s_{11}^{-1}(t)\boldsymbol Q_{1n}(t; \boldsymbol\theta_0, 0) - s_{11}^{-1}(t)s_{12}(t)(\widehat{\boldsymbol\theta} - \boldsymbol\theta_0)+o_p(\delta_n),\quad \text{and}\\
   0 & = & \int^a_{-a}\boldsymbol Q_{2n}(t; \widehat{\boldsymbol\theta}, \boldsymbol\lambda_1(t; \widehat{\boldsymbol\theta}))dG_1(t)\\
   & = &  \int^a_{-a}\left\{\boldsymbol Q_{2n}(t; \boldsymbol\theta_0, 0) + \frac{\partial \boldsymbol Q_{2n}(t; \boldsymbol\theta_0, 0)}{\partial\theta}(\widehat{\boldsymbol\theta}-\boldsymbol\theta_0) + \frac{\partial\boldsymbol Q_{2n}(t; \boldsymbol\theta_0,0)}{\partial\boldsymbol\lambda^\prime}\boldsymbol\lambda_1(t; \widehat{\boldsymbol\theta})\right\}dG_1(t)+o_p(\delta_n)\\
\Rightarrow \quad \widehat{\boldsymbol\theta} - \boldsymbol\theta_0 & = &  -\left\{\int^a_{-a}s_{21}(t)s_{11}^{-1}(t)s_{12}(t)dG_1(t)\right\}^{-1}\times \left\{\int^a_{-a}s_{21}(t)s_{11}^{-1}(t)\boldsymbol Q_{1n}(t; \boldsymbol Q_{1n}(t; \boldsymbol\theta_0, 0))dG_1(t)\right\}\\
   &   & + o_p(\delta_n),
   \end{eqnarray*}
   which completes the proof.
   \end{proof}
   \begin{proof}[Proof of Corollary 5.1]\quad To prove the Corollary, it suffices to show
   \begin{equation*}
   \int^a_{-a}\int^a_{-a}s_{21}(t_1)s_{11}^{-1}(t_1)\boldsymbol\Gamma(t_1, t_2)s_{11}^{-1}(t_2)s_{12}(t_2)dG_1(t_1)dG_1(t_2) - \int^a_{-a}s_{21}(t_1)s^{-1}_{11}(t_1)s_{12}(t_1)dG_1(t_1) \leq 0,
   \end{equation*}
    where $A \leq B$ denotes $A-B$ is a negative-semidefinite. Observe that, by Cauchy-Schwarz inequality, 
   \begin{eqnarray*}
   && \int^a_{-a}\int^a_{-a}s_{21}(t_1)s_{11}^{-1}(t_1)\boldsymbol\Gamma(t_1, t_2)s_{11}^{-1}(t_2)s_{12}(t_2)dG_1(t_1)dG_1(t_2)\\
   & \leq &  \int^a_{-a}\int^a_{-a} \left[s_{21}(t_1)s_{11}^{-1}(t_1)s_{11}(t_1)s_{11}(t_1)s_{21}(t_1)\right]^{1/2}\left[s_{11}(t_2)s_{11}^{-1}(t_2)s_{12}(t_2)s_{11}(t_2)s_{21}(t_2)\right]^{1/2}dG_1(t_1)dG_1(t_2)\\
   & = & \left[\int^a_{-a} \left(s_{21}(t_1)s_{11}^{-1}(t_1)s_{21}(t_1)\right)^{1/2}dG_1(t_1)\right]^{\otimes 2}\\
   & \leq & \int^a_{-a} s_{21}(t_1)s_{11}^{-1}(t_1)s_{21}(t_1)dG_1(t_1).
   \end{eqnarray*}
   It follows that $\boldsymbol\Sigma \leq \left[\int^a_{-a} s_{21}(t)s_{11}^{-1}(t)s_{21}(t)dG_1(t)\right]^{-1}$. The proof can be completed following \cite{Qin_Lawless-1994-AoS}. Write
   \begin{eqnarray*}
   s_{12}(t;\boldsymbol\theta) & = & \left[\left(\frac{\partial \boldsymbol g_1}{\partial\boldsymbol\theta}\right)^\prime, \ldots, \left(\frac{\partial \boldsymbol g_{k-1}}{\partial\boldsymbol\theta}\right)^\prime, \left(\frac{\partial \boldsymbol g_k}{\partial\boldsymbol\theta}\right)^\prime\right] \triangleq \left[s^-_{12}(t;\boldsymbol\theta), \left(\frac{\partial \boldsymbol g_k}{\partial\boldsymbol\theta}\right)^\prime\right]\\
   s_{11}(t; \boldsymbol \theta) & = & \left( \begin{array}{cc} s_{11,a}(t,\boldsymbol\theta) & s_{11,b}(t,\boldsymbol\theta)\\ s_{11,c}(t,\boldsymbol\theta) & s_{11,d}(t,\boldsymbol\theta)\end{array} \right),
   \end{eqnarray*}
   where $s_{11,a}(t, \boldsymbol\theta)$ is a $(k-1)\times(k-1)$ matrix. Then, for all $t\in[-a,a],$
   \begin{eqnarray*}
   s_{21}(t)s^{-1}_{11}(t)s_{21}(t)
   & = & \left[s^-_{12}(t;\boldsymbol\theta), \left(\frac{\partial \boldsymbol g_k}{\partial\boldsymbol\theta}\right)^\prime\right]\left( \begin{array}{cc} s_{11,a}(t,\boldsymbol\theta) & s_{11,b}(t,\boldsymbol\theta)\\ s_{11,c}(t,\boldsymbol\theta) & s_{11,d}(t,\boldsymbol\theta)\end{array} \right)\left[s^-_{12}(t;\boldsymbol\theta), \left(\frac{\partial \boldsymbol g_k}{\partial\boldsymbol\theta}\right)^\prime\right]^\prime\\
   & \geq & \left[s^-_{12}(t;\boldsymbol\theta), \left(\frac{\partial \boldsymbol g_k}{\partial\boldsymbol\theta}\right)^\prime\right]\left( \begin{array}{cc} s_{11,a}(t,\boldsymbol\theta) & \boldsymbol 0\\ \boldsymbol 0 & \boldsymbol 0\end{array} \right)\left[s^-_{12}(t;\boldsymbol\theta), \left(\frac{\partial \boldsymbol g_k}{\partial\boldsymbol\theta}\right)^\prime\right]^\prime\\
   & = & s^-_{12}(t;\boldsymbol\theta) s_{11,a}(t,\boldsymbol\theta)s^-_{12}(t;\boldsymbol\theta)^\prime
   \end{eqnarray*}
   which completes the proof.
   \end{proof}~\\
   \noindent\Large\textbf{Acknowledgment}\normalsize\\~\\
   Wharton Research Data Services (WRDS) was used in preparing the data analysis section presented in this manuscript. This service and the data available thereon constitute valuable intellectual property and trade secrets of WRDS and/or its third-party suppliers. The research is supported by funding from NSF (etc). The reserach was supported in part by grants from NSF-xxxxx, NSF-xxxxx. Tony Sit was supported under Sir Edward Youde Memorial Fellowships (SEYMF) for Overseas Studies.
   \bibliographystyle{asa}
   \bibliography{tsref}
   \newpage
   \begin{table}
   \begin{center}
   \begin{tabular}{|c|c|c|c|c|}
     \hline\hline
            & 0 strike & 1 strike & 2 strikes & 4 strikes \\ \hline
       K    & NA & $0.99S$ & $0.99S$, $1.01S$,  & $0.98$, $0.99S$\\
            & (Chan et al. 2009)& &  & $1.01S$, $1.02S$\\\hline\hline
      $n=125$   & $\widehat{\mu} = 0.044 (0.192)$ & $\widehat{\mu} = 0.056 (0.190)$ & $\widehat{\mu} = 0.045 (0.189)$ & $\widehat{\mu} = 0.041 (0.117)$\\
       & $\widehat{\sigma} = 0.298 (0.019)$& $\widehat{\sigma} = 0.2992 (0.012)$ & $\widehat{\sigma} = 0.2998 (0.009)$ & $\widehat{\sigma} = 0.299 (0.008)$\\\hline
      
      $n=250$   & $\widehat{\mu} = 0.047 (0.133)$ & $\widehat{\mu} = 0.054 (0.132)$ & $\widehat{\mu} = 0.050 (0.130)$ & $\widehat{\mu} = 0.042 (0.130)$  \\
       &  $\widehat{\sigma} = 0.299 (0.014)$ & $\widehat{\sigma} = 0.2996(0.009)$ & $\widehat{\sigma} = 0.2999 (0.0088)$ & $\widehat{\sigma} = 0.2869 (0.073)$\\\hline
      
      $n=500$ &  $\widehat{\mu} = 0.051 (0.097)$ & $\widehat{\mu} = 0.052 (0.091)$ & $\widehat{\mu} = 0.051 (0.0984)$ & $\widehat{\mu} = 0.053
 (0.0873)$\\
       & $\widehat{\sigma} = 0.300 (0.010)$ & $\widehat{\sigma} = 0.2998 (0.0070)$ & $\widehat{\sigma} = 0.2999 (0.0071)$ & $\widehat{\sigma} = 0.2996
 (0.0064)$\\\hline
      
      $n=1000$   & $\widehat{\mu} = 0.047 (0.068)$ & $\widehat{\mu} = 0.054 (0.069)$ & $\widehat{\mu} =  0.054 (0.0069)$ & $\widehat{\mu} = 0.052 (0.0071)$ \\
      & $\widehat{\sigma} = 0.3000 (0.007)$ & $\widehat{\sigma} = 0.2949 (0.0069)$ & $\widehat{\sigma} = 0.295 (0.0066)$ & $\widehat{\sigma} = 0.295 (0.0062)$\\\hline
     \end{tabular}
     \caption{Black-Scholes Model (BS) with true values $(\mu, \sigma) = (0.050, 0.30).$}
     \label{table:simBS}
     \end{center}
     \end{table}
     
     \begin{table}[h]  
     \begin{center}
     \begin{tabular}{|c|c|c|c|c|}
     \hline\hline
            & 0 strike & 1 strike & 2 strikes & 4 strikes \\ \hline
       K    & NA & $0.99S$& $0.99S$, $1.01S$  & $0.98$, $0.99S$\\
            & (Chan et al. 2009)& &  & $1.01S$, $1.02S$\\\hline\hline
      
      $n=125$   & $\widehat{\mu} = 0.068 (0.5829)$ & $\widehat{\mu} = 0.0329 (0.1903)$ & $\widehat{\mu} = 0.0103 (0.1281)$ & $\widehat{\mu} = 0.0244 (0.1644)$ \\
       & $\widehat{\sigma} = 0.1597 (0.3086)$ & $\widehat{\sigma} = 0.2305 (0.1608)$ & $\widehat{\sigma} = 0.2392 (0.1602)$ & $\widehat{\sigma} = 0.2709 (0.1378)$\\
            & $\widehat{\lambda} = 2.7485 (10.0591)$ & $\widehat{\lambda} = 2.8674 (6.0220)$ & $\widehat{\lambda} = 3.1887 (5.4130)$ & $\widehat{\lambda} = 3.0066 (2.8460)$\\
            & $\widehat{\mu_J} = $-0.3121$ (0.6929)$ & $\widehat{\mu_J} = $-0.3445$ (0.4306)$ & $\widehat{\mu_J} = $-0.2749$ (0.2912)$ & $\widehat{\mu_J} = $-0.2670$ (0.2848)$\\
            & $\widehat{\sigma_J} = 0.4440 (0.4946)$ & $\widehat{\sigma_J} = 0.4665 (0.2547)$ & $\widehat{\sigma_J} = 0.4474 (0.2468)$ & $\widehat{\sigma_J} = 0.5009 (0.2273)$\\\hline
            
      
      $n=250$   & $\widehat{\mu} = 0.0639 (0.3790)$ & $\widehat{\mu} = 0.0443 (0.1435)$ & $\widehat{\mu} = 0.0258(0.085)$ & $\widehat{\mu} = 0.0235 (0.0824)$ \\
       & $\widehat{\sigma} = 0.1613(0.2518)$ & $\widehat{\sigma} = 0.2304 (0.1424)$ & $\widehat{\sigma} = 0.2561(0.1303)$ & $\widehat{\sigma} = 0.2806 (0.1250)$\\
                  & $\widehat{\lambda} = 1.8401 (3.5736)$ & $\widehat{\lambda} = 2.3803 (2.9758)$ & $\widehat{\lambda} = 2.5194 (2.0394)$ & $\widehat{\lambda} = 2.5063 (2.0116)$\\
            & $\widehat{\mu_J} = $-0.3172$ (0.4775)$ & $\widehat{\mu_J} = $-0.2993$ (0.3115)$ & $\widehat{\mu_J} = $-0.2811$ (0.2445)$ & $\widehat{\mu_J} = $-0.2723$ (0.2397)$\\
            & $\widehat{\sigma_J} = 0.4229 (0.4559)$ & $\widehat{\sigma_J} = $0.5126$ (0.2216)$ & $\widehat{\sigma_J} = 0.5047 (0.2254)$ & $\widehat{\sigma_J} = 0.5269 (0.2057)$\\\hline
            
      
      $n=500$   & $\widehat{\mu} = 0.0740 (0.2405)$ & $\widehat{\mu} = 0.0650 (0.1228)$ & $\widehat{\mu} = 0.0524 (0.1057)$ & $\widehat{\mu} =  0.0597 (0.1076)$ \\
      & $\widehat{\sigma} =  0.2033 (0.2094)$ & $\widehat{\sigma} = 0.2259 (0.1379)$ & $\widehat{\sigma} = 0.3290 (0.1097)$ & $\widehat{\sigma} = 0.3498 (0.1226)$\\
            & $\widehat{\lambda} = 1.7045 (0.9860)$ & $\widehat{\lambda} = 2.0594 (1.4227)$ & $\widehat{\lambda} = 1.7884 (1.2292)$ & $\widehat{\lambda} = 1.9321 (1.0289)$\\
            & $\widehat{\mu_J} = $-0.2411$ (0.3034)$ & $\widehat{\mu_J} = $-0.2864$ (0.2636)$ & $\widehat{\mu_J} = $-0.3097$ (0.2377)$ & $\widehat{\mu_J} = $-0.3049$ (0.2280)$\\
            & $\widehat{\sigma_J} = 0.5153 (0.2159)$ & $\widehat{\sigma_J} = 0.5203 (0.2143)$ & $\widehat{\sigma_J} = 0.5344 (0.2119)$ & $\widehat{\sigma_J} = 0.5748 (0.1987)$\\\hline
            
      
      $n=1000$   & $\widehat{\mu} = 0.0315 (0.1900)$ & $\widehat{\mu} = 0.0317 (0.0884)$ & $\widehat{\mu} =  0.0326 (0.0825)$ & $\widehat{\mu} = 0.0396 (0.0813)$ \\
     
            & $\widehat{\sigma} = 0.2349 (0.1658)$ & $\widehat{\sigma} = 0.2553 (0.1282)$ & $\widehat{\sigma} = 0.3432 (0.1017)$ & $\widehat{\sigma} = 0.3407 (0.0975)$\\
             & $\widehat{\lambda} = 1.8302 (0.9038)$ & $\widehat{\lambda} = 1.9637 (0.8853)$ & $\widehat{\lambda} = 1.6987 (0.7044)$ & $\widehat{\lambda} = 1.9238 (0.7021)$\\
             & $\widehat{\mu_J} = $-0.2807$ (0.2423)$ & $\widehat{\mu_J} = $-0.3068$ (0.2161)$ & $\widehat{\mu_J} = $-0.2999$ (0.1857)$ & $\widehat{\mu_J} = -0.2776 (0.1627)$\\
             & $\widehat{\sigma_J} = 0.5800 (0.2155)$ & $\widehat{\sigma_J} = 0.6030 (0.1720)$ & $\widehat{\sigma_J} = 0.6095 (0.1618)$ & $\widehat{\sigma_J} = 0.6051 (0.1470)$\\\hline
     \end{tabular}
     \caption{Black-Scholes model with Merton Jumps (BSMJ) with true values $(\mu, \sigma, \lambda, \mu_J, \sigma_J) = (0.0875, 0.30, 2.0, -0.2, 0.60).$ Notice that in this simulation study, we do not follow Chan et al. (2009) true parameter values because it is very unlikely to have jump size with mean and variance $2.0$ and $6.0$ respectively while the underlying drift and variance are mere $0.05\delta$ and $0.30\sqrt{\delta}$.}
     \label{table:simBSMJ}
     \end{center}
     \end{table}

     \begin{table}[h]  
     \begin{center}
     \begin{tabular}{|c|c|c|c|}
     \hline\hline
        & 1 strike & 2 strikes & 4 strikes \\ \hline
       K&  $0.99S$ & $0.99S$   & $0.98S, 0.99S$\\
        &          & $1.01S$   & $1.01S, 1.02S$\\\hline\hline
      
      $n=125$   & $\widehat{\mu} = 0.0974 (0.1840)$ & $\widehat{\mu} = 0.0794 (0.1682)$ & $\widehat{\mu} = 0.0586 (0.1603)$ \\
       & $\widehat{\sigma} = 0.2607 (0.1305)$ & $\widehat{\sigma} = 0.2741 (0.1193)$ & $\widehat{\sigma} = 0.2826 (0.1069)$\\
            & $\widehat{\lambda} = 1.9939 (0.0349)$ & $\widehat{\lambda} = 1.9956 (0.0306)$ & $\widehat{\lambda} = 1.9980 (0.0311)$\\
            & $\widehat{\mu_J} = -0.0874 (0.1070)$ & $\widehat{\mu_J} = -0.1004 (0.0994)$ & $\widehat{\mu_J} = -0.1088 (0.1051)$ \\
            & $\widehat{\sigma_J} = 0.0370 (0.1530)$ & $\widehat{\sigma_J} = 0.0564 (0.1612)$ & $\widehat{\sigma_J} = 0.0592 (0.1642)$\\
            & $\widehat{\alpha} = 0.6956 (1.4105)$ & $\widehat{\alpha} = 0.6460 (1.2068)$ & $\widehat{\alpha} = 0.6429 (1.1079)$ \\\hline
      
      $n=250$   & $\widehat{\mu} = 0.0966 (0.1422)$ & $\widehat{\mu} = 0.0866 (0.1382)$ & $\widehat{\mu} = 0.0732 (0.1315)$ \\
      & $\widehat{\sigma} = 0.2792 (0.1140)$ & $\widehat{\sigma} = 0.2789 (0.1053)$ & $\widehat{\sigma} = 0.2921 (0.0756)$\\
            & $\widehat{\lambda} = 1.9934 (0.0250)$ & $\widehat{\lambda} = 1.9920 (0.0511)$ & $\widehat{\lambda} = 1.9955 (0.0247)$\\
            & $\widehat{\mu_J} = -0.0811 (0.0847)$ & $\widehat{\mu_J} = -0.0927 (0.0818)$ & $\widehat{\mu_J} = -0.1009 (0.0853)$ \\
            & $\widehat{\sigma_J} = 0.0461 (0.1668)$ & $\widehat{\sigma_J} = 0.0496 (0.1659)$ & $\widehat{\sigma_J} = 0.0523 (0.1641)$\\
            & $\widehat{\alpha} = 0.6744 (1.1766)$ & $\widehat{\alpha} = 0.6382 (1.0552)$& $\widehat{\alpha} = 0.6449 (0.9589)$\\\hline
      
      $n=500$   & $\widehat{\mu} = 0.1027 (0.1121)$ & $\widehat{\mu} = 0.0995 (0.1087)$ & $\widehat{\mu} = 0.0876 (0.1047)$ \\
     & $\widehat{\sigma} = 0.2949 (0.0699)$ & $\widehat{\sigma} = 0.2982 (0.0552)$ & $\widehat{\sigma} = 0.2982 (0.0391)$ \\
            & $\widehat{\lambda} = 1.9938 (0.0203)$ & $\widehat{\lambda} = 1.9932 (0.0154)$ & $\widehat{\lambda} = 1.9939 (0.0136)$\\
            & $\widehat{\mu_J} = -0.0841 (0.0693)$ & $\widehat{\mu_J} = -0.0897 (0.0644)$ & $\widehat{\mu_J} = -0.0990 (0.0623)$\\
            & $\widehat{\sigma_J} = 0.0607 (0.1655)$ & $\widehat{\sigma_J} = 0.0605 (0.1635)$ & $\widehat{\sigma_J} = 0.0819 (0.1608)$\\
            & $\widehat{\alpha} = 0.5703 (0.7906)$ & $\widehat{\alpha} = 0.5499 (0.7051)$ & $\widehat{\alpha} = 0.5846 (0.6543)$\\\hline    
      
      $n=1000$  & $\widehat{\mu} = 0.1044 (0.0860)$ & $\widehat{\mu} = 0.1040 (0.0869)$ & $\widehat{\mu} = 0.0918 (0.0855)$ \\
      & $\widehat{\sigma} = 0.3032 (0.0461)$ & $\widehat{\sigma} = 0.3026 (0.0369)$ & $\widehat{\sigma} = 0.3040 (0.0.0279)$\\
            & $\widehat{\lambda} = 1.9946 (0.0096)$ & $\widehat{\lambda} = 1.9947 (0.0080)$ & $\widehat{\lambda} = 1.9960 (0.0068)$\\
            & $\widehat{\mu_J} = -0.0822 (0.0594)$ & $\widehat{\mu_J} = -0.0832 (0.0551)$ & $\widehat{\mu_J} = -0.0946 (0.0494)$ \\
            & $\widehat{\sigma_J} = 0.1007 (0.1488)$ & $\widehat{\sigma_J} = 0.1042 (0.1493)$ & $\widehat{\sigma_J} = 0.1212 (0.1386)$\\
            & $\widehat{\alpha} = 0.5846 (0.6543)$ & $\widehat{\alpha} = 0.5678 (0.6310)$ & $\widehat{\alpha} = 0.5722 (0.5827)$\\\hline
  \end{tabular}
     \caption{Merton Jump-diffusion model with true values $(\mu, \sigma, \lambda, \mu_J, \sigma_J, \alpha) = (0.095, 0.30, 2.0, -0.08, 0.20, 0.60).$}
     \label{table:simMJ-withrisk}
     \end{center}
     \end{table}

         \begin{table}[h]  
     \begin{center}
     \begin{tabular}{|c|c|c|c|}
     \hline\hline
        & 1 strike & 2 strikes & 4 strikes \\ \hline
       K&  $0.99S$ & $0.99S$   & $0.98S, 0.99S$\\
        &          & $1.01S$   & $1.01S, 1.02S$\\\hline\hline
      
      $n=125$   & $\widehat{\mu} = 0.1592 (0.1997)$ & $\widehat{\mu} = 0.1473 (0.1713)$ & $\widehat{\mu} = 0.1129 (0.1777)$ \\
      & $\widehat{\sigma} = 0.2274 (0.1999)$ & $\widehat{\sigma} = 0.3109 (0.0658)$ & $\widehat{\sigma} = 0.3141 (0.0528)$\\
            & $\widehat{\lambda} = 1.7825 (0.9630)$ & $\widehat{\lambda} = 1.8095 (0.6815)$ & $\widehat{\lambda} = 1.9343 (0.6037)$\\
            & $\widehat{p} = 0.1107 (0.7369)$ & $\widehat{p} = 0.2571 (0.9921)$ & $\widehat{p} = 0.3422 (0.5358)$ \\
            & $\widehat{\eta_1} = 7.6935 (0.3580)$ & $\widehat{\eta_1} = 7.6935 (0.3580)$ & $\widehat{\eta_1} = 7.5872 (0.4652)$\\
            & $\widehat{\eta_2} = 8.8874 (0.7513)$ & $\widehat{\eta_2} = 8.9638 (0.5798)$ & $\widehat{\eta_2} = 8.9658 (0.4665)$\\
            & $\widehat{\alpha} = 0.3415 (1.7920)$ & $\widehat{\alpha} = 0.1705 (1.3924)$ & $\widehat{\alpha} = 0.2717 (1.2334)$ \\\hline
      
      $n=250$   & $\widehat{\mu} = 0.1152 (0.1876)$ & $\widehat{\mu} = 0.1115 (0.1527)$ & $\widehat{\mu} = 0.1007 (0.1469)$ \\
      & $\widehat{\sigma} = 0.2969 (0.0641)$ & $\widehat{\sigma} = 0.3018 (0.0443)$ & $\widehat{\sigma} = 0.3023 (0.0606)$\\
            & $\widehat{\lambda} = 1.8130 (0.6496)$ & $\widehat{\lambda} = 1.8397 (0.6693)$ & $\widehat{\lambda} = 1.9636 (0.5531)$\\
            & $\widehat{p} = 0.3424 (0.4645)$ & $\widehat{p} = 0.3934 (0.4164)$ & $\widehat{p} = 0.3897 (0.3653)$ \\
            & $\widehat{\eta_1} = 7.5572 (0.6307)$ & $\widehat{\eta_1} = 7.5784 (0.3063)$ & $\widehat{\eta_1} = 7.5292 (0.4211)$\\
            & $\widehat{\eta_2} = 8.9710 (0.7906)$ & $\widehat{\eta_2} = 9.0307 (0.4292)$ & $\widehat{\eta_2} = 8.9892 (0.4004)$\\
            & $\widehat{\alpha} = 0.1835 (1.2546)$ & $\widehat{\alpha} = 0.0800 (1.1524)$& $\widehat{\alpha} = 0.2256 (1.0019)$\\\hline
      
      $n=500$   & $\widehat{\mu} = 0.1044 (0.1478)$ & $\widehat{\mu} = 0.1206 (0.1239)$ & $\widehat{\mu} = 0.0929 (0.1245)$ \\
      & $\widehat{\sigma} = 0.3065 (0.0355)$ & $\widehat{\sigma} = 0.3036 (0.0329)$ & $\widehat{\sigma} = 0.2976 (0.0487)$ \\
            & $\widehat{\lambda} = 1.8205 (0.6064)$ & $\widehat{\lambda} = 1.8907 (0.7034)$ & $\widehat{\lambda} = 2.0139 (0.5391)$\\
            & $\widehat{p} = 0.4018 (0.3516)$ & $\widehat{p} = 0.4041 (0.3138)$ & $\widehat{p} = 0.4030 (0.3050)$\\
            & $\widehat{\eta_1} = 7.5446 (0.3409)$ & $\widehat{\eta_1} = 7.4794 (0.4696)$ & $\widehat{\eta_1} = 7.4912 (0.3298)$\\
            & $\widehat{\eta_2} = 9.0713 (0.3104)$ & $\widehat{\eta_2} = 9.0671 (0.3318)$ & $\widehat{\eta_2} = 8.9822 (0.2867)$ \\
            & $\widehat{\alpha} = 0.0764 (0.8658)$ & $\widehat{\alpha} = 0.0220 (0.8857)$ & $\widehat{\alpha} = 0.1735 (0.7628)$\\\hline    
      
      $n=1000$  & $\widehat{\mu} = 0.1109 (0.1201)$ & $\widehat{\mu} = 0.1169 (0.1031)$ & $\widehat{\mu} = 0.0725 (0.0955)$ \\
      & $\widehat{\sigma} = 0.2889 (0.0366)$ & $\widehat{\sigma} = 0.2891 (0.0400)$ & $\widehat{\sigma} = 0.2982 (0.0343)$\\
            & $\widehat{\lambda} = 2.0648 (0.6382)$ & $\widehat{\lambda} = 2.1036 (0.7695)$ & $\widehat{\lambda} = 2.0227 (0.4550)$\\
            & $\widehat{p} = 0.4004 (0.2498)$ & $\widehat{p} = 0.4026 (0.2132)$ & $\widehat{p} = 0.4398 (0.2069)$ \\
            & $\widehat{\eta_1} = 7.4740 (0.2350)$ & $\widehat{\eta_1} = 7.4330 (0.3336)$ & $\widehat{\eta_1} = 7.4613 (0.3112)$\\
            & $\widehat{\eta_2} = 8.9613 (0.3358)$ & $\widehat{\eta_2} = 8.9589 (0.3884)$ & $\widehat{\eta_2} = 9.0012 (0.2128)$\\
            & $\widehat{\alpha} = 0.1242 (0.6115)$ & $\widehat{\alpha} = 0.0254 (0.5691)$ & $\widehat{\alpha} = 0.1828 (0.4250)$\\\hline
  \end{tabular}
     \caption{Kou Double-exponential Jump-diffusion model with true values $(\mu, \sigma, \lambda, p, \eta_1, \eta_2, \alpha) = (0.095, 0.30, 2.0, 0.05, 7.5, 9.0, 0.20).$}
     \label{table:simDEJD}
     \end{center}
     \end{table}

     \begin{table}[h]  
     \begin{center}
     \begin{tabular}{|c|c|c|c|}
     \hline
            0 strike                       & 1 strike                            & 2 strikes                      & 4 strikes \\
            Chan et al. (2009)&&&\\ \hline
            $\widehat{\mu} = 0.0620 (0.0296)$    & $\widehat{\mu} = -3.5483\times 10^{-10} (0.0179)$ & $\widehat{\mu} = 0.0133 (0.0030)$    & $\widehat{\mu} = 0.0133 (7.7640\times 10^{-6})$ \\
            $\widehat{\sigma} = 0.1327 (0.0036)$ & $\widehat{\sigma} = 0.2058 (0.0025)$      & $\widehat{\sigma} = 0.2257 (0.0019)$ & $\widehat{\sigma} = 0.2340 (6.2043 \times 10^{-4})$\\\hline
     \end{tabular}
\label{tab:BSDA}  
     \caption{Empirical estimation for the S\&P500 index between January 2, 1987 and September 29, 2008: Black-Scholes Model.}
     \end{center}
     \end{table}
   
     \begin{table}[h]  
     \begin{center}
     \label{tab:BSMJDA}
     \begin{tabular}{|c|c|c|c|}
     \hline
            0 strike						& 1 strike							& 2 strikes							& 4 strikes \\ 
            Chan et al. (2009)&&&\\ \hline
            $\widehat{\mu} = 0.0800 (0.0296)$		& $\widehat{\mu} = -0.0763 (0.0031)$	& $\widehat{\mu} = -0.0731 (0.0009)$	& $\widehat{\mu} = -0.0755 (0.0005)$ \\
            $\widehat{\sigma} = 0.12715 (0.0137)$	& $\widehat{\sigma} = 0.2101 (0.0177)$	& $\widehat{\sigma} = 0.2374 (0.0080)$	& $\widehat{\sigma} = 0.2379 (0.0048)$\\
            $\widehat{\lambda} = 1.24319 (0.0049)$	& $\widehat{\lambda} = 1.8688 (0.0002)$ 	& $\widehat{\lambda} = 1.8760 (0.0011)$	& $\widehat{\lambda} = 1.876 (0.0001)$\\
            $\widehat{\mu_J} = 0.0213 (2.3519)$	& $\widehat{\mu_J} = 0.0177 (0.9440)$	& $\widehat{\mu_J} = -0.0050 (0.0622)$	& $\widehat{\mu_J} = -0.0061 (0.0283)$\\
            $\widehat{\sigma_J} = 0.0267 (2.5951)$	& $\widehat{\sigma_J} = 0.0821 (0.943)$& $\widehat{\sigma_J} = 0.0223 (0.2606)$	& $\widehat{\sigma_J} = 0.0226 (0.1270)$\\
            NA							& $\widehat{\alpha} = 0.2860 (2.2553)$  	& $\widehat{\alpha} = 0.2889 (1.5988)$	& $\widehat{\alpha} = 0.2934 (1.3815)$\\\hline
     \end{tabular}
     \caption{Empirical estimation for the S\&P500 index between January 2, 1987 and September 29, 2008: Black-Scholes with Merton Jumps Model.}
     \end{center}
     \end{table}

     \begin{table}[h]  
     \begin{center}
     \label{tab:DEJDDA}
     \begin{tabular}{|c|c|c|c|}
     \hline
            0 strike						& 1 strike						& 2 strikes						& 4 strikes \\ 
            Chan et al. (2009)&&&\\ \hline
            $\widehat{\mu} = 0.0606 (0.0279)$		& $\widehat{\mu} = -0.0438 (0.0173)$		& $\widehat{\mu} = 0.0035 (0.0034)$		& $\widehat{\mu} = -0.0775 (0.0014)$ \\
            $\widehat{\sigma} = 0.0924 (0.0048)$	& $\widehat{\sigma} = 0.09447 (0.0043)$	& $\widehat{\sigma} = 0.1956 (0.0014)$	& $\widehat{\sigma} = 0.2246 (0.0014)$\\
            $\widehat{\lambda} = 2.160 (0.1067)$	& $\widehat{\lambda} = 2.1493 (0.0828)$	& $\widehat{\lambda} = 2.1684 (0.0371)$	& $\widehat{\lambda} = 1.6002 (0.0323)$\\
            $\widehat{p} = 0.2715 (0.0001)$		& $\widehat{p} = 0.2408 (0.0001)$		& $\widehat{p} = 0.4576 (0.0001)$		& $\widehat{p} = 0.4633 (0.0001)$\\
            $\widehat{\eta_1} = 16.1389 (0.2696)$	& $\widehat{\eta_1} = 16.1391 (0.2761)$	& $\widehat{\eta_1} = 16.1434 (0.0775)$	& $\widehat{\eta_1} = 21.5180 (0.0589)$\\
            $\widehat{\eta_2} = 25.2204 (1.6940)$	& $\widehat{\eta_2} = 25.2201 (1.5885)$	& $\widehat{\eta_2} = 20.1736 (0.5765)$	& $\widehat{\eta_2} = 30.2644 (1.3122)$\\     
            NA	& $\widehat{\alpha} = 0.09093 (7.9389)$	& $\widehat{\alpha} = 0.2084 (0.4855)$	& $\widehat{\alpha} = 0.2329 (0.1077)$\\\hline
     \end{tabular}
     \caption{Empirical estimation for the S\&P500 index between January 2, 1987 and September 29, 2008: Kou Double Exponential Jump-Diffusion Model.}
     \end{center}
     \end{table}

\end{document}